\documentclass[%
 reprint,
 amsmath,amssymb,
 aps,
]{revtex4-2}
\usepackage{graphicx}
\usepackage{dcolumn}
\usepackage{bm}
\begin{document}

\preprint{APS/123-QED}

\title{Ripple Signatures of Majorana Hybridization across a Topological Quantum Quench}
	
	\author{Xin-Xin Wang}
	\affiliation{School of Physics, Henan Normal University, Xinxiang 453007, Henan Province, China}
	\author{Jin-Xin Li}
	\affiliation{School of Physics, Henan Normal University, Xinxiang 453007, Henan Province, China}
    \author{Ya-Wen Tang}
	\affiliation{School of Physics, Henan Normal University, Xinxiang 453007, Henan Province, China}
    \author{Lu Qin}
	\affiliation{School of Physics, Henan Normal University, Xinxiang 453007, Henan Province, China}
    \author{Zun-Lue Zhu}
	\affiliation{School of Physics, Henan Normal University, Xinxiang 453007, Henan Province, China}
    \author{Wu-Ming Liu}
	\affiliation{School of Physics, Henan Normal University, Xinxiang 453007, Henan Province, China}
    \author{Xing-Dong Zhao}
	\email{phyzhxd@gmail.com}
	\affiliation{School of Physics, Henan Normal University, Xinxiang 453007, Henan Province, China}
	\author{Liang-Liang Wang}
	\email{wangliangliang@westlake.edu.cn}
	\affiliation{School of Science, Westlake University, 600 Dunyu Road, Hangzhou 310030, Zhejiang Province, China}
	\affiliation{Institute of Natural Sciences, Westlake Institute for Advanced Study, 18 Shilongshan Road, Hangzhou 310024, Zhejiang Province, China}

\date{\today}

\begin{abstract}
    The crossover between topology and nonequilibrium dynamics has emerged as a rich frontier, in which quantum systems can exhibit unique dynamical phenomena that lie beyond the reach of equilibrium.
    Of particular interest are quench dynamics across topological phase, as it may reveal the information about the underlying Majorana zero-energy states. 
    Here, we investigate the fate of Majorana boundary modes in a quenched fermionic superfluid using self-consistent time-dependent Bogoliubov-de Gennes theory.
    For quantum quenches within the topological regime, Majorana boundary modes survive but undergo coherent boundary oscillations arising from the nonadiabatic deformation of their wave functions. 
    Furthermore, a pronounced ripple pattern appears in the post-quench density distribution following a sudden topology-changing quench.
    Here we identify that these ripple structures originates from the coherent hybridization and interference of the initially separated Majorana boundary states.
    Our findings establish nonequilibrium boundary dynamics as a new approach for probing Majorana physics, which is complementary to conventional equilibrium measurements.
	\end{abstract}
	
	\maketitle

	\section{Introduction}
	Topological properties of quantum many-body systems have been a subject of intense research, owing to both their fundamental physical significance and their potential applications in fault-tolerant topological quantum computation\cite{AdyStern,RevModPhys.80.1083,3,4,Eisert2015}. 
    Because topological quantum operations often require controlled switching between trivial and nontrivial phases\cite{PhysRevB.97.060304,Dong2015,PhysRevE.90.032138,Chung2016},
    understanding the resulting quantum dynamics is therefore essential not only for identifying nonequilibrium topological phenomena, but also for assessing the speed, fidelity, and limitations of topological-state manipulation\cite{DAlessio2015,RevModPhys.80.885,PhysRevLett.111.053003}.
    The precise tunability of interactions, geometry, and driving protocols makes ultracold atomic gases an ideal platform for time-resolved studies of nonequilibrium dynamics that remain difficult to access in solid-state systems\cite{Trotzky2012,scienceMeinert,RevModPhys.80.885,PhysRevLett.111.053003}. 
    In the past decade, $s$-wave Bardeen-Cooper-Schrieffer (BCS) superfluid with spin-orbit coupling (SOC) have attracted significant attention: combined with SOC, Zeeman field and $s$-wave attractive interactions, such systems can lead to effective $p$-wave pairing which supports topological states\cite{PhysRevA.85.033622,PhysRevLett.109.095301,PhysRevLett.109.115301,Liu2014, PhysRevLett.96.097005,PhysRevLett.103.020401,Liao2010,34,35,37,Fan2022,RevModPhys.83.863}.

	Among many unique properties of topological phases, of particular importance is the presence of Majorana zero modes (MZMs), that often give rise to robust encoding of quantum information against local perturbations\cite{Qu2013,Hegde_2015,Wang_2015,Wilczek2009,mourik2012signatures,PhysRevLett.105.077001,mourik2012signatures,PhysRevLett.105.077001,name2, Lutchyn2018}. 
    They obey non-Abelian exchange statistics and their antiparticles are themselves, satisfying $\gamma(z)=\gamma^{\dagger}(z)$, can exhibit distinct properties unlike conventional fermions, thus have been proposed to be building blocks of fault-tolerant quantum computation\cite{PhysRevLett.105.077001,Chung2016}.
    Most previous studies have focused on equilibrium detection or bulk post-quench observables, whereas the real-time evolution of localized Majorana modes remains less understood.
    In particular, static topological protection guarantees the existence of Majorana boundary modes, but does not ensure that an initially prepared Majoranas remains dynamically unchanged under rapid quenching.
    This raises the central question of what is the role of topological properties in the quench dynamics? 
    What observable physical phenomena arise from the post-quench dynamics of Majorana zero modes? 
    While there exist some reports on such studies, a comprehensive understanding has yet to emerge, especially when Majorana states are involved.

	In this work, we investigate the nonequilibrium evolution of Majorana boundary modes in a one-dimensional spin-orbit-coupled fermionic gas subjected to Zeeman field quenches \cite{Delfino_2014}.
    Using self-consistent time-dependent Bogoliubov-de Gennes simulations, we construct a boundary-sensitive dynamical phase diagram that extends earlier classifications based primarily on the bulk pairing order parameter \cite{Wang_2015,Dong2015}.
    For quenches confined to the topological regime, the post-quench Hamiltonian continues to support Majorana modes, while the mismatch between the initial and final quasi-particle bases produces coherent boundary oscillations whose amplitude increases with the degree of non-adiabaticity.
    In contrast, a quench from the topological to the trivial phase removes the zero-mode protection and projects the initially separated Majorana components onto finite-energy Bogoliubov states. 
    Their inward propagation, hybridization, and interference generate pronounced spatiotemporal ripple structures in the density response.
    Complementary finite-time ramps exhibit the same qualitative mechanism (see Appendix \ref{app:B} for more informations).
    Our results identify time-resolved boundary dynamics as a probe of Majorana boundary modes beyond conventional bulk observables.
	
	The structure of this paper is as follows. In Sec.~\ref{Sec: model} we first introduce the microscopic model and the self-consistent time-dependent Bogoliubov--de Gennes framework used to describe the nonequilibrium evolution of the spin-orbit-coupled Fermi superfluid. 
    In Sect.~\ref{sec:result}-A, we then construct a boundary-sensitive dynamical phase diagram by quenching the Zeeman field across different regions of the equilibrium phase diagram.
    The resulting dynamics reveal two distinct boundary responses: coherent Majorana oscillations for quenches within the topological regime (Sect. \ref{sec:result}-B), and interference-driven ripple dynamics for quenches crossing the topological transition (Sect. \ref{sec:result}-C). 
    By analyzing the evolution of the lowest-energy quasiparticle wave functions and their particle-hole self-conjugacy fidelity, we identify the mechanisms governing the survival, hybridization, and eventual loss of Majorana character. 
    Finally, we discuss the implications of these results for probing nonequilibrium topological states through time-resolved density measurements and for understanding the limitations of static topological protection under dynamical driving in Sec.~\ref{Sec: conclusion}.

\section{Model and Quench Protocol}
\label{Sec: model}
In this study, the quench dynamics of a one-dimensional spin-orbit coupled fermionic gas with attractive $s$-wave interactions are considered.
The Hamiltonian is written as $\hat{H}=\hat{H}_{0}+\hat{H}_{\mathrm{int}}$, where 
\begin{equation}
\begin{split}
\label{Eq:H_single}
 \hat{H}_{0}&= \int dx \hat{\Psi}^\dagger\left[ -\frac{\hbar^2}{2m}\partial_x^2 + V_{\mathrm{trap}} -\mu- i\alpha \hbar\partial_x \sigma_y + h_z \sigma_z \right] \hat{\Psi},\\
\hat{H}_{\mathrm{int}}&=\int dx g_{\mathrm{1D}}\hat{\psi}^{\dagger}_{\uparrow}(x)\hat{\psi}^{\dagger}_{\downarrow}(x)\hat{\psi}_{\downarrow}(x)\hat{\psi}_{\uparrow}(x).\\
 \end{split}
\end{equation}
Here $\hat{\Psi}\equiv [\hat{\psi}_{\uparrow}(x),\hat{\psi}_{\downarrow}(x)]^{T}$ are annihilation operators for fermions with spin-up and spin-down.
$V_{\mathrm{trap}}(x)=m\omega^2x^2/2$ is the trap potential, with $\mu$ being the chemical potential, $\alpha$ is the SOC strength and $h_z$ denotes the effective Zeeman field.
The Zeeman term breaks time-reversal symmetry and, together with spin-orbit coupling and attractive pairing, realizes an effective spinless $p$-wave channel capable of supporting Majorana boundary modes.
Within the standard mean-field approximation, we introduce the pairing order parameter $ \Delta(x,t) = -g_{\mathrm{1D}} \langle \hat{\psi}_\downarrow(x) \hat{\psi}_\uparrow(x) \rangle $, then the interaction is decoupled as $\hat{H}_{\text{int}} \rightarrow \Delta(x,t) \hat{\psi}_\uparrow^\dagger(x) \hat{\psi}_\downarrow^\dagger(x)+ \Delta^\ast(x,t) \hat{\psi}_\downarrow(x) \hat{\psi}_\uparrow(x) - |\Delta(x,t)|^2/g_{\mathrm{1D}}$.
Under the Nambu basis, $\Psi_\eta(x) = [u_{\uparrow,\eta}(x), u_{\downarrow,\eta}(x), v_{\downarrow,\eta}(x), -v_{\uparrow,\eta}(x)]^\mathrm{T}$, the quasiparticle dynamics are governed by the time-dependent Bogoliubov--de Gennes (TDBdG) equation:
\begin{equation}
\label{eq:H_bdg}
i\hbar\frac{\partial}{\partial t}\Psi_\eta(x,t)=
\begin{pmatrix}
H_{0} & \Delta(x,t) \\
\Delta^{\ast}(x,t) & -\sigma_{y}H_{0}^{\ast}\sigma_{y} \\
\end{pmatrix}
\Psi_\eta(x,t),
\end{equation}
with the pairing order parameter is updated self-consistently according to 
\begin{equation}
\begin{split}
\Delta(x, t)=g_{\mathrm{1D}} \sum_{E_{\eta}>0}[ &u_{\uparrow,\eta}(x,t)v_{\downarrow,\eta}^{\ast}(x,t) f(-E_\eta)\\
&+u_{\downarrow,\eta}(x,t)v_{\uparrow,\eta}^{\ast}(x,t) f(E_\eta)],\\
\end{split}
\end{equation}
with $f(E) = [\exp({E/\mathrm{k}_{B} T}) + 1]^{-1}$ being the Fermi-Dirac distribution at temperature $T$. 
The strength of the interaction is governed by the effective one-dimensional coupling constant $g_{\mathrm{1D}} = -2\hbar^2/(m a_{\mathrm{1D}})$, where $a_{\mathrm{1D}}$ is the $s$-wave scattering length.
The spin-resolved density is
\begin{equation}
\label{density-self-consistent}
n_\sigma(x, t) = \sum_{E_{\eta}>0} \left[ |v_{\sigma,\eta}(x, t)|^2 f(-E_\eta) + |u_{\sigma,\eta}(x, t)|^2 f(E_\eta) \right].
\end{equation}
Eqs. (\ref{eq:H_bdg})-(\ref{density-self-consistent}) provide a self-consistent description of both the collective superfluiding condensate and individual quasiparticle evolutions\cite{PhysRevLett.93.160401}. 
The nonequilibrium protocol is implemented by suddenly changing the Zeeman field:
$h_z(t)=h_i+\left(h_f-h_i\right)\Theta(t)$,
where $\Theta(t)$ denotes the Heaviside step function.
For $t<0$, the system is initially prepared in the self-consistent ground state of the Hamiltonian with Zeeman field $h_i$. 
At $t=0$, the Zeeman field is abruptly switched to $h_f$.
The post-quench Hamiltonian is then held fixed while the many-body state evolves according to the TDBdG equation. This protocol injects energy non-adiabatically and projects the initial quasiparticle wave functions onto the eigenmodes of the final Hamiltonian.
It therefore provides a direct way to distinguish the survival of the topological boundary sector from the dynamical stability of a particular Majorana wavefunction profile.

In the simulations, we adopt dimensionless units where the Fermi wave vector $k_F = \pi n / 2$ and the Fermi energy $E_{F} = \hbar^2 k_F^2 / (2m)$ serve as the reference scales. The interaction strength is characterized by the parameter $\gamma = -m g_{1D} / (\hbar^2 n)$. We consider a system with total atom number $N = 100$ and use the peak density of a noninteracting Fermi gas in the Thomas--Fermi approximation, $n = (2/\pi)\sqrt{N m \omega / \hbar}$, as the reference density. All simulations are performed at zero temperature with a time step of $dt=10^{-4}T_{F}$, chosen to balance computational feasibility with numerical accuracy.

\section{Results and Discussion }\label{sec:result}
\subsection{Quench Phase Diagram}\label{sec:result-a}
Before analyzing the nonequilibrium dynamics, we first recall the static topological structure of the system. 
In a one-dimensional spin-orbit-coupled fermionic superfluid, Zeeman field plays a dual role: it breaks time-reversal symmetry and drives the system through the topological phase transition. 
Within the mean-field framework, the transition is signaled by the bulk pairing gap closing at the critical field
$
h_c=\sqrt{\mu^2+\Delta^2}.
$
For $h_z<h_c$, the system is topologically trivial and possesses a fully gapped bulk spectrum\cite{PhysRevLett.93.160401,PhysRevLett.96.097005}. 
When $h_z$ exceeds $h_c$, the gap closes and subsequently reopens with a different topological character, leading to the emergence of topological protected Majorana zero modes localized near the system boundaries \cite{PhysRevLett.102.085302,Zhang2013,PhysRevLett.126.193401,HaibinWu, fidelity1}. 

Quantum Quenches provides a natural way to probe how these topological features respond away from equilibrium. 
Previous studies of quenched spin-orbit-coupled fermionic gases has primarily classified the dynamical states through the long-time behavior of the pairing order parameter. 
Ref.~\cite{Dong2015} identified three dynamical regimes: an undamped oscillatory phase, a damped oscillatory phase approaching a finite plateau, and an overdamped phase in which the pairing order parameter decays toward zero. 
This classification captures the collective bulk response after a quench and has provided an important framework for understanding nonequilibrium superfluid dynamics.
However, such a bulk-centered description does not fully address the fate of topological boundary modes. 
Majorana zero modes are spatially localized, particle-hole self-conjugate quasiparticles whose stability is tied to the topology of the bulk Hamiltonian, but their real-time evolution after a sudden parameter change need not be determined solely by the long-time behavior of $\Delta(t)$. 
In particular, a quench can project the initial Majorana wave functions onto a superposition of low-energy boundary states and finite-energy Bogoliubov quasiparticles of the post-quench Hamiltonian. 
This projection may lead to boundary oscillations, hybridization with bulk Bogoliubov quasiparticles, or loss of Majorana self-conjugacy.
Therefore, a complete description of topological quench dynamics requires not only the tracking of the bulk order parameter, but also the spatially resolved evolution of the boundary-localized quasiparticle modes.
 \begin{figure}[t]
	\centering  
	\begin{minipage}{\columnwidth}
		\centering
		\includegraphics[width=\columnwidth]{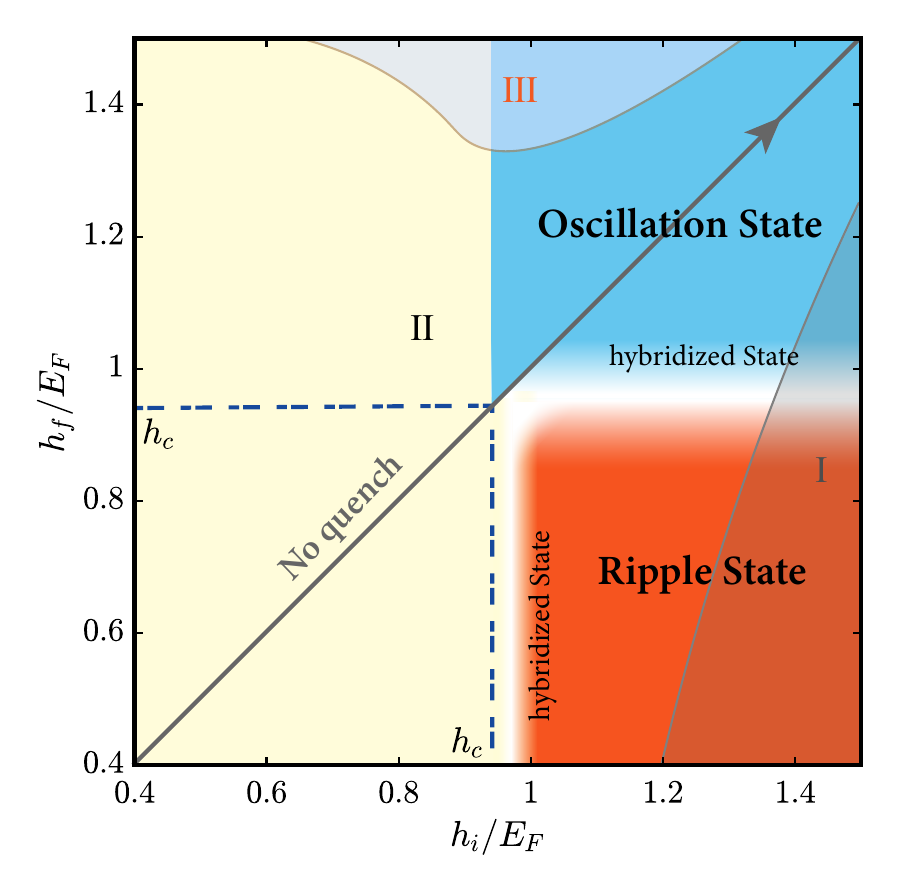}
		\caption{(Color online) Quench dynamical phase diagram. The regions are determined 
by the time-evolved density distribution after a Zeeman-field quench from $h_i$ to $h_f$. The diagonal line $h_i = h_f$ corresponds to no quench, where the system remains in its initial state. The equilibrium critical point $h_c = 0.94E_F$ separates the topological and trivial phases.  According to the long-time asymptotic behavior of the order parameter, the phase diagram is first partitioned into three regions (I, II, and III, dashed lines); from these, based on the dynamical behavior of the boundary states, we distill two fundamental states. Two dynamical phases are shaded in blue (oscillation state) and orange (ripple state). The white gradient area indicates that the system has entered the hybridized state. In the ripple phase, the density evolution exhibits continuous oscillations with pronounced ripple features. In the oscillation phase, the time evolution of the density distribution at the boundaries exhibits pronounced oscillatory behavior. }
		\label{fig:1}
	\end{minipage}
\end{figure}

\begin{figure*}[t]
	\centering  
	\begin{minipage}{\textwidth}
		\centering
		\includegraphics[width=\textwidth]{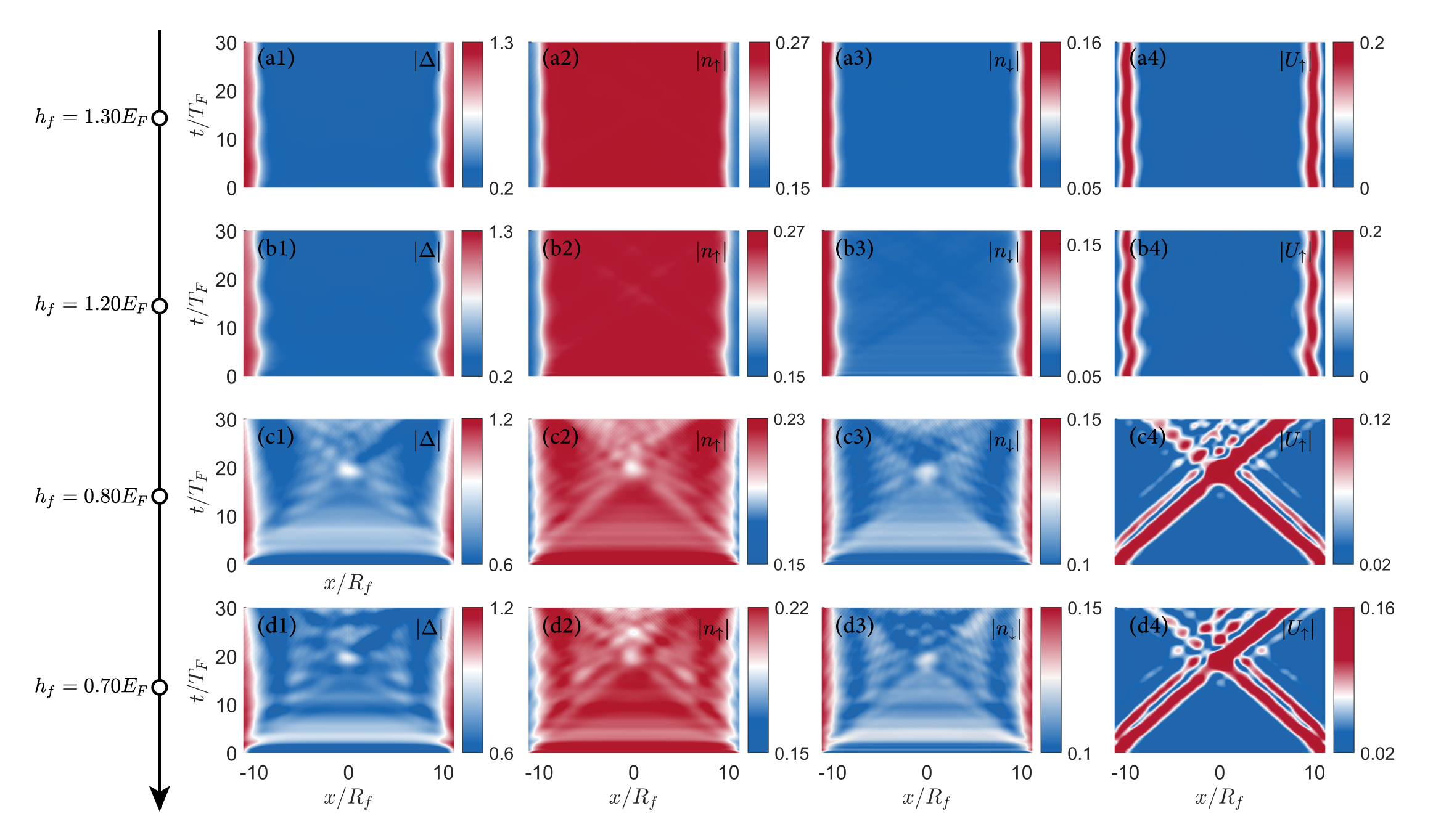}
		\caption{(Color online) The figure shows the sequence parameters and density distribution after quenching, as well as the time evolution of the lowest-energy wavefunctions. When the final Zeeman field \(h_f\) is tuned such that the quench path crosses the quantum phase transition point, the dynamical response of the system exhibits a distinct ripple phase. This feature stands in marked contrast to the case where \(h_f\) does not cross the transition point. The quenched Majorana zero mode in the topological region exhibits localized oscillatory behavior. All results are obtained for a fixed initial field \(h_i = 1.40E_{F}\). }
		\label{fig:delta+density} 
	\end{minipage}
\end{figure*}

In this work, we introduce a spatially potential that allows boundary states to be engineered within a prescribed region of the system. 
This construction enables a boundary-sensitive dynamical phase diagram from the density response (shown in Fig.~\ref{fig:1}), in the parameter space spanned by $h_i$ and $h_f$.
For quenches confined in the topological phase, $h_i,h_f>h_c$, the quench preserves the existence of Majorana boundary modes.
However, the initial Majorana wave functions are not eigenstates of the post-quench Hamiltonian and their projection onto the final low-energy and quasiparticle states produces coherent boundary oscillations in the local density, defining the oscillation phase. 
This regime should be viewed as a nonadiabatic deformation fo surviving Majorana modes.
A qualitatively different response occurs for quenches from the topological phase into the trivial phase.
The initially separated Majorana components acquire finite- energy character, propagate toward the trap center, and hybridize with each other and with bulk Bogoliubov excitations. The resulting interference generates pronounced ripple structures in the density evolution, which we identify as the ripple phase. 
This mechanism is absent for quenches inside the trivial phase or for reverse quenches from the trivial to the topological phase, emphasizing that the observed response is controlled not only by the final Hamiltonian but also by the topological content of the initial state.
The white regions in Fig.~\ref{fig:1} correspond to hybridized regimes where the dynamics cannot be cleanly classified as either oscillatory or ripple-like.  
The elaborate evolution associated with these regimes will be analyzed in detail below.

\subsection{Quench-Induced Majorana Oscillation within the Topological Region}
\label{sec:result-oscillation}

\begin{figure}[t]
\centering
\includegraphics[width=\columnwidth]{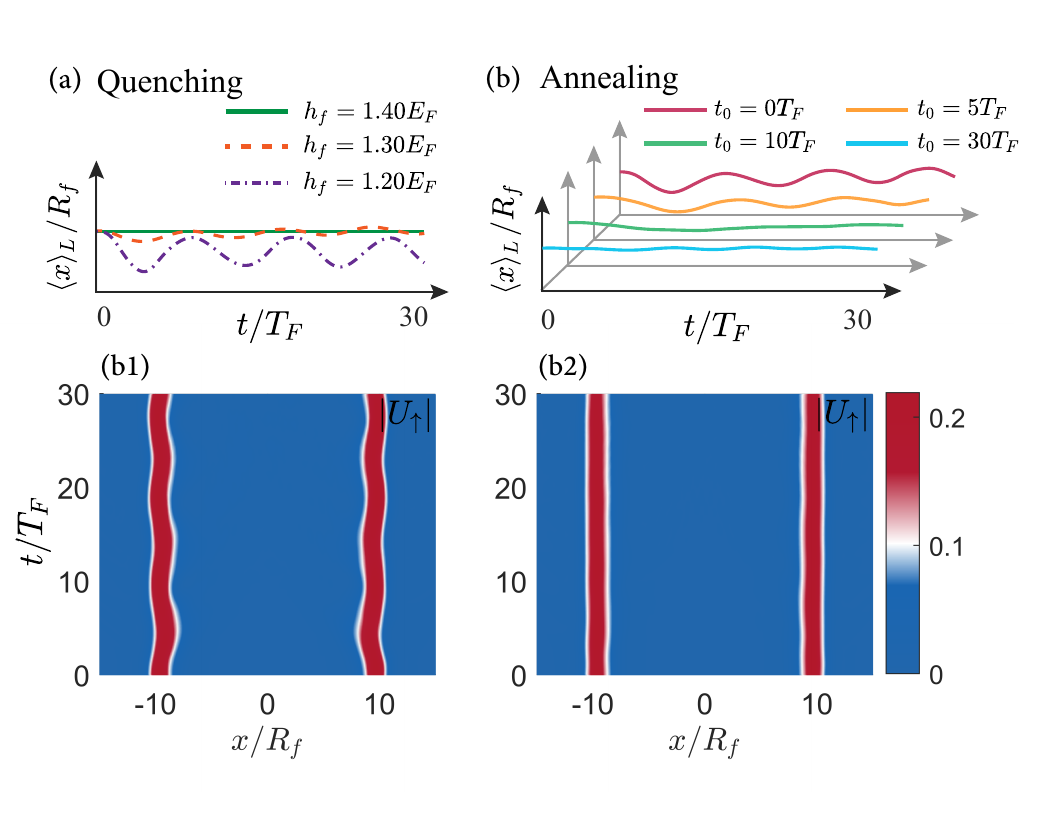}
\caption{(Color online) Majorana wave-packet dynamics driven by a quantum quench. 
(a) For a fixed initial Zeeman field $h_i = 1.40E_{F}$, increasing the quench amplitude renders the coherent Majorana zero-mode oscillations more pronounced.
(b) Wave-packet dynamics under a linear ramp described by $H_z(t) = h_i + \frac{h_f - h_i}{t_0} 
\bigl[ t \,\Theta(t) - (t - t_0) \,\Theta(t - t_0) \bigr]$, where the Zeeman field is ramped from $h_i = 1.30E_{F}$ to $h_f = 1.15E_{F}$. As the annealing time $t_0$ increases, the oscillations gradually weaken.
(b1), (b2) Space-time evolution of the lowest-energy wave function for $t_0 = 0\,T_{F}$ (sudden quench) and $t_0 = 30\,T_{F}$ (slow annealing), respectively.
}
\label{fig:3}
\end{figure}

We first examine the quenches within the topological phase. 
Along this protocol, the Hamiltonian is changed but the bulk topological invariant is not.
The post-quench Hamiltonian continues to support Majorana boundary modes, but the initial Majorana wave functions are generally no longer eigenmodes of the final Hamiltonian\cite{Chung2016} .
Thus the quenching dynamics should be viewed as the coherent evolution of an initially prepared Majorana boundary state projected onto the renewed Majorana section of the post-quench system.
Their subsequent phase evolution produces coherent boundary dynamics rather than an equilibrium reconstruction of the final Majorana mode.
Figures~\ref{fig:delta+density}(a, b) show the spatiotemporal evolution of the pairing order parameter $\Delta(x,t)$, the spin-resolved density $n_{\sigma}(x,t)$, and the corresponding lowest-energy quasiparticle wave functions within the topological phase. 
The system is initialized at $h_i=1.4E_{F}$, where a pair of Majorana zero modes is localized near the boundaries, and the final Zeeman field $h_f$ is varied while remaining in the topological regime, such as $h_{f}=1.3E_{F}$ in \ref{fig:delta+density}(a) and $h_{f}=1.2E_{F}$ in \ref{fig:delta+density}(b).
As shown in Fig. \ref{fig:delta+density}, both $\Delta(x,t)$ and $n_{\sigma}(x,t)$ develop pronounced boundary-localized oscillations, which is qualitatively different from the bulk collective dynamics reported in Ref.~\cite{Dong2015}. 
In those works, the leading diagnostic is the spatially extended time dependence of the order parameter, often associated with collective amplitude dynamics of the condensate and approximately symmetric with respect to the trap center.
Here the strongest signal is instead concentrated near
the system edges and is correlated with the lowest-energy
quasiparticle wave functions. 
These oscillations show that the Majorana sector survives the perturbation, but its wave-function profile is dynamically dressed by the mismatch between the initial and final Bogoliubov bases.
Fig. \ref{fig:3}(a) further illustrates how the boundary response depends on the quench strength.
As $|h_f-h_i|$ increases, the boundary oscillation grows and extend itself further into the bulk. 
The physical boundary itself is fixed, this apparent inward shift should not be interpreted as motion of the boundary or of a topological defect. 
It reflects a redistribution and deformation of the boundary-localized quasiparticle weight under nonadiabatic evolution \cite{Wang2021,PhysRevLett.91.250402}.

We next adobe finite-time annealing to separate Majorana-sector stability from nonadiabatic excitation effects. 
The Zeeman field is varied linearly from $h_i=1.3E_{F}$ to $h_f=1.15E_{F}$, with the entire protocol remaining inside the topological phase. 
Figure~\ref{fig:3}(b) shows the lowest-energy wave-packet dynamics for several annealing durations $t_0=5,10,30\,T_{F}$, while Figs.~\ref{fig:3}(c) and \ref{fig:3}(d) compare the sudden-quench limit with the slow-annealing case $t_0=30T_{F}$.
Here we note that, as $t_0$ increases, the boundary oscillations are progressively suppressed, which is consistent with a crossover from nonadiabatic dynamics to approximate adiabatic one. 
Thus, the oscillations are not intrinsic equilibrium features of the final Majorana mode, but nonequilibrium signatures of nonadiabatic driving.

\subsection{Ripple Phase Induced by a Quench Across the Topological Transition}
\label{sec:result-collisions}

\begin{figure}[t]
\centering
\begin{minipage}{\columnwidth}
\centering
\includegraphics[width=\columnwidth]{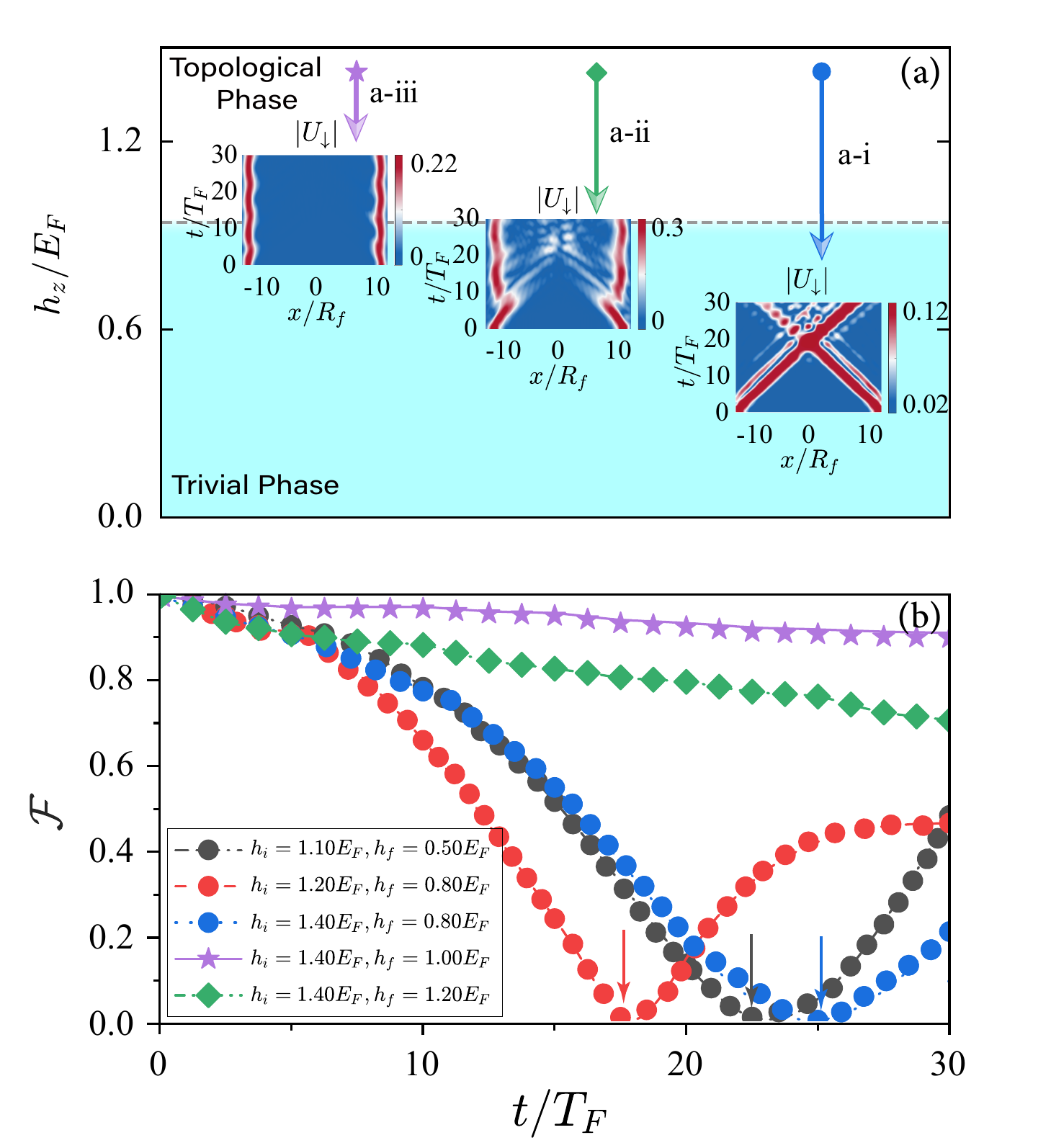}
\caption{(Color online) (a) Post-quench dynamics of the lowest-energy Bogoliubov quasiparticle mode. The dashed line marks the quantum critical field $h_c$ separating the topological (T) and trivial (Tr) phases. The arrows indicate the Zeeman-field quench protocols considered in panels (a-i)--(a-iii);  the associated parameters are given in panel (b). 
(b) Time-dependent Majorana fidelity $\mathcal{F}$ for the different quench protocols. Quenches across the topological transition (red, black, and blue curves) produce a rapid decrease of $\mathcal{F}(t)$ toward zero, indicating the loss of Majorana self-conjugacy caused by strong hybridization of the two boundary components. For the quench confined to the topological phase (purple curve), the fidelity remains high, demonstrating the persistence of the edge-localized Majorana character. For the near-critical quench (green curve), the fidelity decreases below $0.8$ but does not collapse abruptly to zero, consistent with an intermediate regime involving substantial coupling to low-energy bulk excitations.}
\label{fig:4}
\end{minipage}
\end{figure}

We now turn to the quenches that cross the topological phase, with the system initially prepared in the topological phase and suddenly changed to a final field away from the critical field $h_f<h_c$. This protocol differs qualitatively from the above intra-topological quench: the final Hamiltonian no longer supports protected zero-energy boundary eigenstates and the initially localized Majoranas will projected onto a superposition of finite-energy Bogoliubov quasiparticles of the post-quench Hamiltonian\cite{PhysRevB.96.125113,Dziarmaga01112010}.
Consequently, the resulting dynamics exhibit strong spatiotemporal modulations in both the pairing field $\Delta(x,t)$ and the spin-resolved densities $n_\sigma(x,t)$:
both the pairing order parameter $\Delta(x,t)$ and the spin-resolved densities $n_\sigma(x,t)$ develop pronounced spatial fluctuations and exhibits a characteristic ripple pattern, as shown in Figs.~\ref{fig:delta+density}(c) and \ref{fig:delta+density}(d).
For the initial prepared Majorana modes, we observe that they propagate inward and gradually hybridize, and finally collide with each others, inducing a pronounced peak in the pairing order parameter and density distribution at the center. 
Here we note that the ripple pattern is closely related to the inward propagation, interference, and eventual annihilate of the initially separated Majorana components.
This quench dynamics structure can be measured as a direct probe of Majorana zero modes and provide a dynamical signature that is complementary to interferometry and tunneling spectroscopy.
The situation is distinct when the final field is around the critical field.
As shown in Fig.~\ref{fig:4}(a-ii), the bulk excitation gap becomes small, which introduces a long dynamical timescale and enhances the coupling between the edge modes and low-energy bulk excitations. The lowest-energy wave function consequently retains part of its initial edge-localized structure for a finite time, but later develops strong spatial fluctuations. In this regime, neither a stable boundary Majorana mode nor a complete central annihilate can be clearly identified.

Here we introduce the normalized particle--hole self-conjugacy overlap, which we refer to as the Majorana fidelity $\mathcal{F}$ to quantify the dynamical loss of Majorana character\cite{PhysRevB.99.155419,fidelity2,fidelity3,fidelity4,fidelity5,fidelity6}. 
For the lowest-energy Bogoliubov quasiparticle, $u_{\sigma}(x,t)$ and $v_{\sigma}(x,t)$ denote the particle- and hole- components. 
We define
\begin{align}
\mathcal{F}(t)
&= \frac{\left| \left\langle v(x,t) \middle| u(x,t) \right\rangle \right|}{|u|\cdot|v|} \nonumber \\
&= \frac{
    \left|
      \displaystyle \sum_{\sigma} \int u_\sigma(x,t) v_\sigma(x,t) \, dx
    \right|
  }{
    \sqrt{
      \displaystyle \sum_{\sigma} \int |u_\sigma(x,t)|^2 \, dx
    }
    \sqrt{
      \displaystyle \sum_{\sigma} \int |v_\sigma(x,t)|^2 \, dx
    }
  }
  \label{eq:f}
\end{align}
This quantity follows the Majorana self-conjugate condition $u_{\sigma}(x)=v_{\sigma}^{\ast}(x)$ with the Cauchy--Schwarz inequality $\mathcal{F}\in[0,1]$.
Values close to unity indicate a well-defined self-conjugate Majorana-like state, while a reduced value measures hybridization with finite-energy quasiparticles.
A value close to unity indicates that the particle component $u_{\sigma}$ and the particle--hole-conjugated hole component $v_{\sigma}^{\ast}$ coincide up to a global phase, as expected for a well-defined Majorana zero mode. By contrast, a reduced value signals increasing deviation from self-conjugacy and stronger hybridization with finite-energy quasiparticle states \cite{UHLMANN,Jozsa01121994,nielsen2010quantum,WANG200858}. The numerical values $0.99$ and $0.8$ used in the following discussion serve only as practical indicators for the present calculations and should not be regarded as universal thresholds.


The fidelity dynamics in Fig.~\ref{fig:4}(b) provide a quantitative counterpart to the spatial evolution in Fig.~\ref{fig:4}(a).  
For quenches from the topological phase into the trivial phase, $\mathcal F$ remains large only during the short interval in which the wave packets still retain memory of their initial boundary localization\cite{Chung2016,Lee2021}.  
As the two components move inward and overlap, $\mathcal F$ rapidly decreases toward zero, signaling that the lowest-energy quasiparticle no longer satisfies the Majorana self-conjugacy condition.  
This collapse identifies the ripple phase with the dynamical conversion of separated Majorana components into finite-energy Bogoliubov excitations.
For the intratopological quench, $h_i=1.40E_{F}$ and $h_f=1.20E_{F}$, the fidelity remains high throughout the simulated time window, confirming that the Majorana sector is dynamically robust when the post-quench Hamiltonian stays topological.  
For the near-critical quench, $h_f=1.00E_{F}$, the fidelity decreases gradually but does not collapse immediately.  
This intermediate behavior reflects the small critical gap, which enhances coupling between the edge sector and low-energy bulk modes while preserving partial Majorana character.  
Thus, the fidelity tracks the full degradation pathway of Majorana zero modes and distinguishes coherent boundary oscillation, boundary-bulk hybridization, and complete loss of self-conjugacy.  
More broadly, this analysis shows that topological quench dynamics cannot be inferred solely from the final Hamiltonian: the observable response also depends on the topological content of the initial state, the quench direction, finite-size splitting, trap inhomogeneity, and the available bulk quasiparticle spectrum.

\section{Conclusion}
\label{Sec: conclusion}
We have investigated the nonequilibrium dynamics of Majorana boundary modes in a quenched one-dimensional spin-orbit-coupled Fermi superfluid.
For quenches confined within the topological phase, the Majorana sector
survives.
Nevertheless, a sudden parameter change projects the initial
Majorana wave functions onto the eigenstates of the post-quench Hamiltonian,
producing coherent oscillations localized near the boundaries.
These oscillations represent a dynamical deformation of protected boundary
states rather than a breakdown of topological protection.
A qualitatively different response occurs when the quench crosses the
topological transition. 
The initially separated Majorana components retain their boundary localization only transiently before propagating toward the center and hybridizing with each other and with finite-energy Bogoliubov
quasiparticles. 
Their coherent interference generates pronounced ripple
structures in the density and order-parameter evolution. 
The ripple phase therefore provides a direct dynamical signature of the loss
of Majorana character after a topology-changing quench.
Near the critical point, the vanishing bulk excitation gap enhances coupling
between boundary modes and bulk quasiparticles, producing an intermediate
hybrid regime. 
To quantify the dynamical evolution of Majorana character, we introduce a
particle-hole self-conjugacy fidelity. 
This quantity distinguishes persistent Majorana dynamics from hybridization
with finite-energy excitations and provides a quantitative measure of
boundary-mode stability beyond conventional bulk probes.

Our results establish a connection between topology, boundary-state dynamics,
and experimentally accessible density responses. 
The predicted ripple structures and Majorana fidelity evolution provide
possible signatures for detecting and manipulating nonequilibrium Majorana
physics in ultracold atomic platforms. 
More broadly, this work demonstrates that topology constrains the existence
of boundary states but does not uniquely determine their real-time dynamics,
highlighting the importance of boundary-sensitive observables in quantum
quench experiments.

\begin{acknowledgments}
	This work was supported by the National Natural Science Foundation of China (62505079, 12404377), the China Postdoctoral Science Foundation (2025M783380), the Key International Cooperation Project in Henan Province (261111521200), the Natural Science Foundation of Henan Province (252300421995) and the Training Program for Young Back Bone Teachers in Higher Education Institutions of Henan Province (2024GGJS046).
\end{acknowledgments}

\appendix 
\section*{Appendix: Mean-field formulation and supplementary quench analysis}
This appendix provides the theoretical framework used to describe a one-dimensional spin-orbit-coupled Fermi gas with attractive $s$-wave interactions, which serves as the basis for our analysis of topological superconductivity and Majorana zero modes.

\section{Mean-field Hamiltonian and time-dependent BdG equations}
\label{app:A}

We consider a one-dimensional spin-orbit-coupled fermionic gas with two hyperfine spin states, denoted by $\uparrow$ and $\downarrow$. The many-body Hamiltonian is written as
\begin{equation}
\hat H=\hat H_0+\hat H_{\mathrm{int}},
\end{equation}
where
\begin{equation}
\begin{split}
\label{app:Eq:H_single}
 \hat{H}_{0}&= \int dx \hat{\Psi}^\dagger\left[ -\frac{\hbar^2}{2m}\partial_x^2 + V_{\mathrm{trap}} -\mu- i\alpha \hbar\partial_x \sigma_y + h_z \sigma_z \right] \hat{\Psi},\\
\hat{H}_{\mathrm{int}}&=\int dx g_{\mathrm{1D}}\hat{\psi}^{\dagger}_{\uparrow}(x)\hat{\psi}^{\dagger}_{\downarrow}(x)\hat{\psi}_{\downarrow}(x)\hat{\psi}_{\uparrow}(x).\\
 \end{split}
\end{equation}
Here$\hat\Psi(x)=
\left[
\hat\psi_\uparrow(x),\hat\psi_\downarrow(x)
\right]^T$ is the fermionic field spinor. The field operators satisfy
\begin{equation}
\begin{split}
\{\hat\psi_\sigma(x),\hat\psi_{\sigma'}^\dagger(x')\}
&=\delta_{\sigma\sigma'}\delta(x-x'),\\
\{\hat\psi_\sigma(x),\hat\psi_{\sigma'}(x')\}
&=0.
\end{split}
\end{equation}
The trap potential is denoted by $V_{\mathrm{trap}}(x)$, $\mu$ is the chemical potential, $\alpha$ is the spin-orbit-coupling strength, and $h_z$ is the effective Zeeman field. The Zeeman term breaks time-reversal symmetry. In combination with spin-orbit coupling and attractive pairing, it generates an effective spinless $p$-wave channel that can support Majorana boundary modes. The one-dimensional interaction strength is parameterized by
$g_{\mathrm{1D}} = -2\hbar^2/(m a_{\mathrm{1D}})$,
where $a_{\mathrm{1D}}$ is the effective one-dimensional scattering length.

Within the mean-field approximation, we introduce the pairing order parameter
$\Delta(x,t)=-g_{\mathrm{1D}}
\left\langle
\hat\psi_\downarrow(x,t)\hat\psi_\uparrow(x,t)
\right\rangle.$
The interaction Hamiltonian is then decoupled as
$\hat{H}_{\text{int}} \rightarrow \Delta(x,t) \hat{\psi}_\uparrow^\dagger(x) \hat{\psi}_\downarrow^\dagger(x)+ \Delta^\ast(x,t) \hat{\psi}_\downarrow(x) \hat{\psi}_\uparrow(x) - |\Delta(x,t)|^2/g_{\mathrm{1D}}$.
In the Nambu representation, the quasiparticle wave function is expressed as $\Psi_\eta(x) = [u_{\uparrow,\eta}(x), u_{\downarrow,\eta}(x), v_{\downarrow,\eta}(x), -v_{\uparrow,\eta}(x)]^\mathrm{T}$, where $\eta$ labels the quasiparticle state. The quasiparticle wave functions obey the time-dependent Bogoliubov--de Gennes equation
\begin{equation}
i\hbar\frac{\partial}{\partial t}\Psi_\eta(x,t)
=
\mathcal H_{\mathrm{BdG}}(x,t)\Psi_\eta(x,t),
\label{app:eq:TDBdG}
\end{equation}
with
\begin{equation}
\mathcal H_{\mathrm{BdG}}(x,t)=
\begin{pmatrix}
H_0(x) & \Delta(x,t)\\
\Delta^\ast(x,t) & -\sigma_y H_0^\ast(x)\sigma_y
\end{pmatrix},
\label{app:eq:BdGBlock}
\end{equation}
the BdG Hamiltonian can be written explicitly as
\begin{equation}
\mathcal H_{\mathrm{BdG}}=
\begin{pmatrix}
\mathcal K+h_z & -\mathcal D & \Delta & 0\\
\mathcal D & \mathcal K-h_z & 0 & \Delta\\
\Delta^\ast & 0 & -\mathcal K+h_z & \mathcal D\\
0 & \Delta^\ast & -\mathcal D & -\mathcal K-h_z
\end{pmatrix}.
\label{app:eq:BdGExplicit}
\end{equation}
where $\mathcal K(x)=-\frac{\hbar^2}{2m}\partial_x^2+V_{\mathrm{trap}}(x)-\mu,
\mathcal D=\alpha\hbar\partial_x$.

The quasiparticle wave functions are used to update the order parameter self-consistently:
\begin{equation}
\begin{split}
\Delta(x,t)=g_{\mathrm{1D}}
\sum_{E_\eta>0}
\biggl[
&u_{\uparrow,\eta}(x,t)v^\ast_{\downarrow,\eta}(x,t)f(-E_\eta)\\
&+u_{\downarrow,\eta}(x,t)v^\ast_{\uparrow,\eta}(x,t)f(E_\eta)
\biggr],
\end{split}
\label{app:eq:gap}
\end{equation}
where$f(E)=1/[{e^{E/(k_B T)}+1}]$ is the Fermi--Dirac distribution. At zero temperature, $f(E>0)=0$ and $f(-E)=1$, so Eq.~\eqref{app:eq:gap} reduces to
\begin{equation}
\Delta(x,t)=g_{\mathrm{1D}}
\sum_{E_\eta>0}
 u_{\uparrow,\eta}(x,t)v^\ast_{\downarrow,\eta}(x,t).
\end{equation}
The spin-resolved density is
\begin{equation}
n_\sigma(x,t)=
\sum_{E_\eta>0}
\left[
|v_{\sigma,\eta}(x,t)|^2f(-E_\eta)
+|u_{\sigma,\eta}(x,t)|^2f(E_\eta)
\right].
\label{app:eq:density}
\end{equation}
The total density is given by
\begin{equation}
n(x,t)=n_\uparrow(x,t)+n_\downarrow(x,t),
\end{equation}

For a homogeneous system with $V_{\mathrm{trap}}=0$ and a constant pairing field $\Delta$, the quasiparticle spectrum can be obtained analytically by replacing $\partial_x\rightarrow ik$. With
\begin{equation}
\xi_k=\frac{\hbar^2 k^2}{2m}-\mu,
\qquad
E_k=\sqrt{\xi_k^2+|\Delta|^2},
\end{equation}
the two positive BdG branches are
\begin{equation}
E_\pm(k)=
\sqrt{
E_k^2+h_z^2+\hbar^2\alpha^2 k^2
\pm
2\sqrt{
 h_z^2E_k^2+\xi_k^2\hbar^2\alpha^2 k^2
}
}.
\label{app:eq:spectrum}
\end{equation}
The bulk gap closes at $k=0$ when
\begin{equation}
h_z^2=\mu^2+|\Delta|^2.
\label{app:eq:topcond}
\end{equation}
Thus, the homogeneous topological criterion is
\begin{equation}
h_z^2>\mu^2+|\Delta|^2.
\end{equation}
In a trapped system, this criterion should be understood as a useful local and qualitative indicator; the actual topological boundary modes must be identified from the finite-system BdG spectrum and the spatial profile of the corresponding low-energy states.

\section{Supplementary analysis of quench paths and finite-time annealing protocols}
\label{app:B}

This section supplements the main discussion by comparing three representative nonequilibrium protocols: a finite-time ramp from the topological phase to the trivial phase, a sudden quench confined entirely within the trivial phase, and a sudden quench from the trivial phase to the topological phase. These protocols clarify which post-quench oscillations are generic bulk quasiparticle responses and which features are directly associated with the low-energy boundary sector.

To examine the effect of a finite ramp rate, we consider a linear annealing protocol for the Zeeman field,
\begin{equation}
h_z(t)=
\begin{cases}
h_i, & t<0,\\[2pt]
h_i+(h_f-h_i)t/t_0, & 0\leq t\leq t_0,\\[2pt]
h_f, & t>t_0,
\end{cases}
\label{app:eq:ramp}
\end{equation}
where $t_0$ is the total ramp duration. In the nearly adiabatic limit, $t_0$ should be long compared with the inverse characteristic many-body energy scale; in practice, ramp times on the order of or longer than $1/E_F$ are typically required to suppress strong nonadiabatic excitations~\cite{Wang_2015,Fan2022}. 

\begin{figure}[t]
	\centering  
	\begin{minipage}{\columnwidth}
		\centering
    \includegraphics[width=\columnwidth]{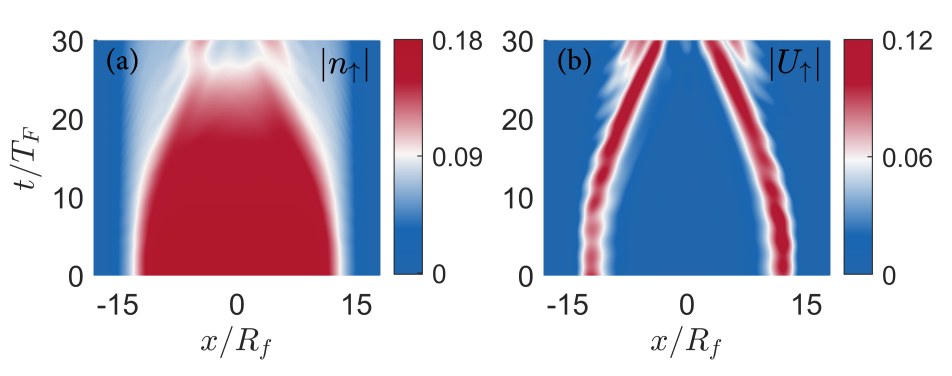}
		\caption{Dynamics of a finite-time ramp from the topological phase to the trivial phase. The Zeeman field is linearly ramped from $h_i=1.00E_F$ to $h_f=0.80E_F$ over a total time $T_F$, crossing the critical value $h_c$ at approximately $t=0.9T_F$ measured from the beginning of the ramp. Panels (a) and (b) show the time evolution of the spin density $n_\uparrow(x,t)$ and the post-ramp quasiparticle wave-function profile, respectively.}
		\label{fig:anneal}
	\end{minipage}
\end{figure}
Unlike a sudden quench, a sufficiently slow annealing protocol allows the system to follow the instantaneous low-energy state more closely. However, when the annealing path crosses the topological critical region, the reduction of the bulk excitation gap inevitably enhances nonadiabatic transitions. As a result, pronounced ripple structures can still emerge even under a controlled finite-time ramp, as shown in Fig.~\ref{fig:anneal}.
The appearance of these ripples is qualitatively consistent with the cross-critical quench dynamics presented in Fig.~\ref{fig:4}. In both cases, the time-dependent variation of the Zeeman field excites low-energy bulk quasiparticle modes near the critical point. The propagation and interference of these excitations produce oscillatory spatial structures in the order parameter and density distributions. The ripple pattern should therefore be interpreted as a characteristic nonequilibrium response to the closing of the excitation gap and the accompanying change in the topology of the instantaneous Hamiltonian.
\begin{figure}[t]
    \centering
    \begin{minipage}{\columnwidth}
        \centering
        \includegraphics[width=\columnwidth]{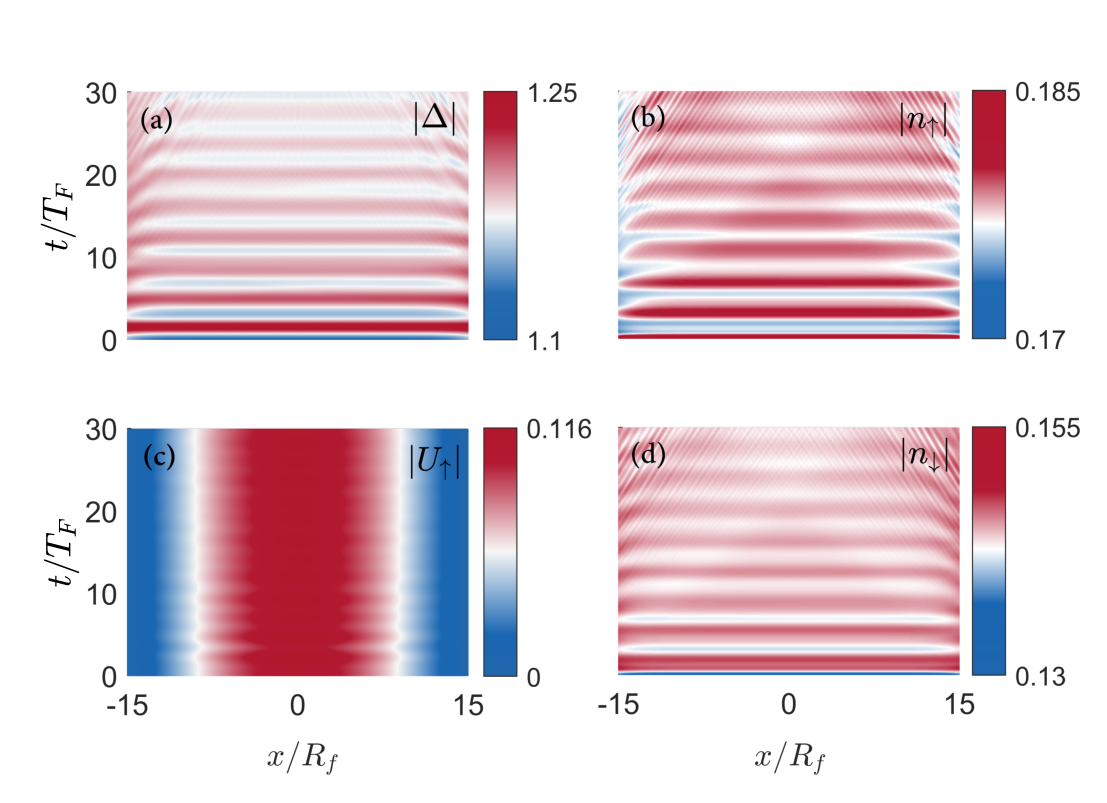}
        \caption{(Color online) Quench dynamics confined to the topologically trivial phase. The Zeeman field is suddenly changed from $h_i=0.70E_{F}$ to $h_f=0.50E_{F}$, with both fields satisfying $h_i,h_f<h_c$. The pairing order parameter, spin-resolved density, and lowest-energy quasiparticle wave function all exhibit coherent post-quench oscillations.}
        \label{fig:topo-off}
    \end{minipage}
\end{figure}
Fig.~\ref{fig:anneal}(b) further shows that the two initially separated Majorana components move toward the trap center as the system is driven from the topological phase into the trivial phase. As their spatial overlap increases, the two components hybridize and acquire a finite-energy splitting. They therefore cease to exist as independent zero-energy boundary modes. This process is more accurately described as the annihilation and hybridization.
After the ramp, the final Hamiltonian lies in the topologically trivial phase and no longer supports protected zero-energy boundary modes. In a closed system, the post-ramp state is generally not identical to the ground state of the final Hamiltonian because residual quasiparticle excitations remain. Thus, the system does not automatically relax to the trivial ground state under unitary evolution. Instead, it evolves as a coherent superposition of eigenstates of the final Hamiltonian. 

To determine whether oscillatory behavior is unique to Majorana modes, we also consider a sudden quench performed entirely within the topologically trivial phase. As shown in Fig.~\ref{fig:topo-off}, the Zeeman field is changed from $h_i=0.70E_{F}$ to $h_f=0.50E_{F}$ with both $h_i$ and $h_f$ below the critical field $h_c$.
Neither the initial nor the final Hamiltonian supports Majorana zero modes, the order parameter, density distribution, and lowest-energy quasiparticle wave function still exhibit pronounced coherent oscillations. 
However, it is only bulk oscillation rather than boundary oscillation.
This initial state can be expressed as a coherent superposition of a series of eigenstates of the new Hamiltonian (including collective excitation modes and quasiparticle excitations), thereby giving rise to coherent oscillations of physical quantities 
such as the order parameter. Subsequently, various decoherence and dissipation mechanisms progressively destroy the quantum coherence, leading to the suppression of the oscillations. Eventually, the system relaxes to a steady state (e.g., a generalized Gibbs ensemble or a thermal state), whose properties are determined by the dissipation strength, the integrability of the system itself, and the details of the quench. 
Oscillations occurring within the trivial phase originate from generic interference among finite-energy quasiparticle modes. By contrast, for a quench confined to the topological phase, the dominant oscillatory structures remain localized near the boundaries and are correlated with the dynamics of the Majorana-related low-energy modes. The spatial localization, quasiparticle energy, and particle--hole self-conjugacy must therefore be examined together to identify the Majorana contribution to the dynamics.
\begin{figure}[t]
    \centering
    \begin{minipage}{\columnwidth}
        \centering
        \includegraphics[width=\columnwidth]{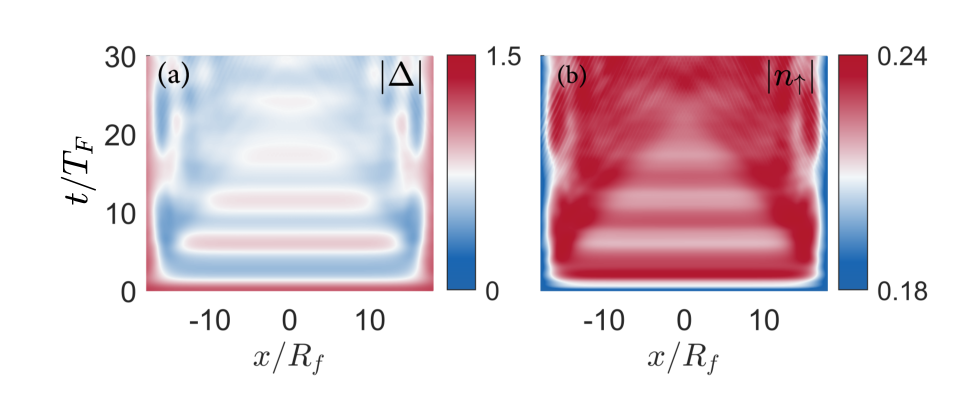}
        \caption{(Color online) Quench dynamics from the trivial phase to the topological phase. The Zeeman field is suddenly changed from $h_i=0.60E_{F}$ to $h_f=1.10E_{F}$. The post-quench evolution of the pairing order parameter $\Delta(x,t)$ and the spin-resolved density $n_\sigma(x,t)$ is shown.}
        \label{fig:8}
    \end{minipage}
\end{figure}

We next consider a quench from the topologically trivial phase into the topological phase($h_i=0.60E_{F}, h_f=1.10E_{F}$).
Before the quench, the initial Hamiltonian does not support Majorana zero modes, and its lowest-energy quasiparticle state generally has a finite excitation energy and is not required to be localized at the boundaries. After the quench, however, the final Hamiltonian lies in the topological regime and supports low-energy boundary modes.
Due to the projection of the initial finite-energy quasiparticle state onto both the boundary and bulk eigenmodes of the final topological Hamiltonian. 
These components possess different energies and spatial profiles and therefore accumulate different dynamical phases during the evolution. Their interference produces the observed propagation, deformation, and gradual redistribution of the wave-function density.

\nocite{*}

\end{document}